\documentclass[aps,prd,floatfix,superscriptaddress,a4paper,
               nofootinbib,notitlepage,12pt]{revtex4-2}

\usepackage[colorlinks=true,linkcolor=blue,citecolor=blue,urlcolor=blue]{hyperref}

\usepackage{graphicx}
\usepackage{amsmath}
\usepackage{amssymb}
\newcommand{\hsp}{\hspace*{1pt}}

\pdfoutput=1

\begin{document}

\title{Why are the dilepton temperatures 
at~relativistic heavy-ion~colliders constant, $T \sim 0.3~\mathrm{GeV}$?}

\author{H. Stoecker}

\affiliation{
Frankfurt Institute for Advanced Studies, Frankfurt am Main, Germany}

\affiliation{
Institut f\"ur Theoretische Physik, J.~W.~Goethe Universit\"at,
Frankfurt am Main, Germany}

\affiliation{
GSI Helmholtzzentrum f\"ur Schwerionenforschung GmbH, Darmstadt, Germany}

\author{L.~M.~Satarov}
\affiliation{
Frankfurt Institute for Advanced Studies, Frankfurt am Main, Germany}

\author{V.~Vovchenko}
\affiliation{
Department of Physics, University of Houston, USA}

\begin{abstract}
The STAR collaboration at RHIC and the ALICE collaboration at the LHC
have reported dielectron spectra in the intermediate mass region,
\mbox{$M_{e^+e^-} = (1-3)$} GeV, which reveal a strikingly constant,
energy-independent emission temperature 
$T_{IMR} \simeq 0.3~\textrm{GeV}$
over a broad range of collision energies,
$\sqrt{s_{NN}} = 27 - 5020~\textrm{GeV}$.
This unexpected ``thermostat'' behavior raises fundamental questions:
why does the temperature remain constant despite increasing collision energy,
and what mechanism governs this apparent universality?
\end{abstract}

\flushbottom
\maketitle

\thispagestyle{empty}

\section*{Introduction}

Invariant mass spectra of dileptons, which barely interact with gluons, 
quarks and hadrons,  can serve as a unique thermometer for the hottest initial stage of 
relativistic nuclear collisions. The NA60 collaboration at CERN's SPS accelerator  
pioneered these temperature measurements with dimuon mass spectra from fixed target In+In 
experiments~\cite{NA60,NA60:2010muk} at $\sqrt {s_{NN}}=17.3~\textrm{GeV}$.

The recent article~\cite{STAR:2024bpc} of the STAR Collaboration
reports the measurements of intermediate mass dilepton spectra
from Au+Au collisions at
$\sqrt {s_{NN}} = 27$ and $54.4~\textrm{GeV}$ at BNL's RHIC collider.
Here, we complement these results by STAR's data~\cite{STAR15,STAR24} for 
RHIC's top energy $\sqrt{s_{NN}}=200~\textrm{GeV}$. Furthermore, the ALICE Collaboration
at the CERN LHC reported recently~\cite{ALICE24,ALICE25} measurements of dielectron spectra  
for Pb+Pb collisions at $\sqrt{s_{NN}} = 5.02~\textrm{TeV}$. For all currently published 
energies, from $\sqrt {s_{NN}} = 27~\textrm{GeV}$ to $5020~\textrm{GeV}$ (an increase
of a~factor 200 in c.m. bombarding energy), the temperatures $T_{IMR}$ extracted from the 
slopes of intermediate mass dilepton spectra equal approximately $0.3~\textrm{GeV}$.
The four STAR data alone yield an average value \mbox{$T_{IMR}\simeq 287\pm 27~\textrm{MeV}$} 
(see Fig.~\ref{fig1}).     

All these measured dilepton temperatures are remarkably close to the critical
temperature \mbox{$T_c^{\hsp Y\hspace*{-2pt}M}\sim 290~\textrm{MeV}$~\cite{Bor22}}
of the first-order phase transition obtained in the pure SU(3) gauge theory.
According to Refs.~\cite{Fujimoto:2025sxx,cgyz-3956}, the latter
is slightly below the gluonic Hagedorn temperature, which
characterizes the melting of glueballs (closed gluon strings) in pure
gauge matter.
If the earliest stage is chemically undersaturated in quarks, then the apparent 
dilepton IMR temperature may become sensitive not to the highest initial gluon temperature, 
but to the stage when quarks become sufficiently abundant, possibly 
near $T_c^{\hsp Y\hspace*{-2pt}M}$.

\section*{Effective temperatures of intermediate mass dileptons 
observed at different energies from SPS to LHC}

\begin{figure}[htb!]
\centering
\includegraphics[trim=2cm 7.5cm 3cm 8.5cm,width=0.65\textwidth]{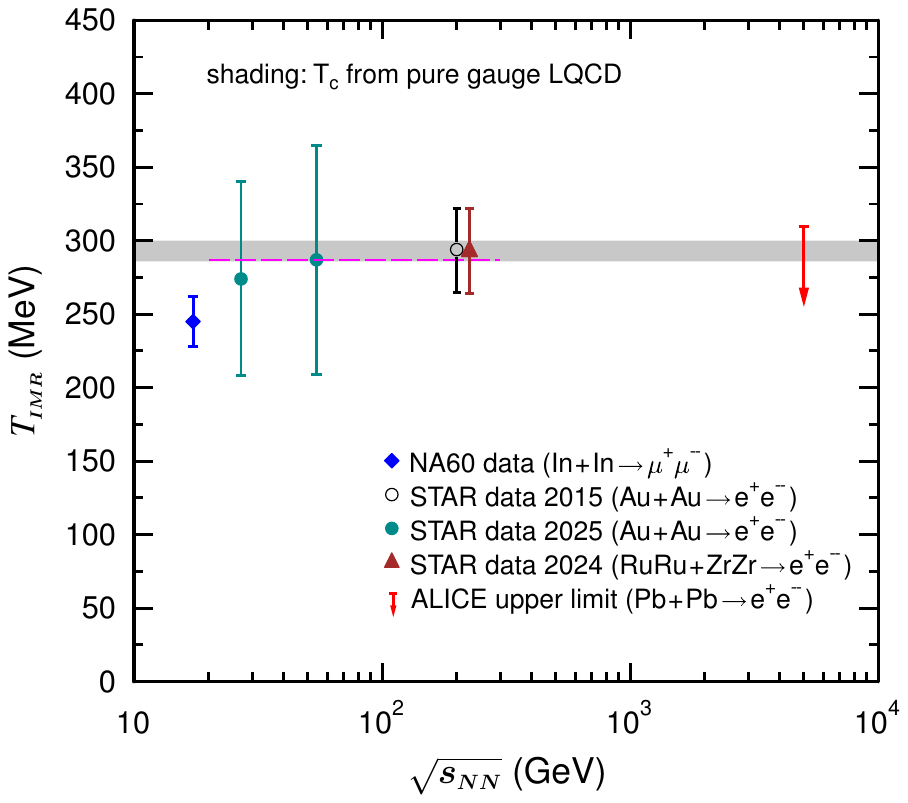}
\caption{
The five experimentally measured temperatures of IMR
dielectrons are shown. Here IMR denotes the intermediate mass region,
$1~\textrm{GeV}<M<3~\textrm{GeV}$. For clarity, the STAR 2024 point is shifted
to the right from its actual position by 25 GeV along the horizontal axis.
The observed temperatures are constant
within $\sim 10\%$ for all nuclear collisions' experimental data
 at the various bombarding energies at BNL and CERN.
 The theoretically determined temperature of the deconfinement phase
 transition of pure gluon matter (see text) is shown by the shaded
 horizontal band. The dashed
line marks the temperature, obtained by arithmetic averaging of the
four experimentally determined dilepton temperatures from STAR (RHIC). The
arrow indicates the upper temperature limit extracted from the 
ALICE data~\cite{ALICE24,ALICE25} at~CERN.
}\label{fig1}
\end{figure}

Figure~1 shows the $T_{IMR}$ values extracted from momentum integrated
dilepton spectra
at various bombarding energies from SPS~\cite{Mynote1}\nocite{NA60} to LHC. 
These temperatures
were determined from fits to the thermal formula for dilepton mass
distributions of the form~\cite{Rap16}
$dN_{l^+l^-}/dM\propto M^{3/2}\exp{(-M/T_{IMR})}$,
where $M$ is the invariant mass of the dilepton pair.

The fits to the dilepton mass spectra in the IMR use the measured
data obtained by the experimental collaborations after subtracting the
so-called 'cocktail' contributions of final hadron decays
(including mesons with $c, \overline{c}, b, \overline{b}$ quarks),
as well as Drell-Yan emission from initial nucleons.
Possible additional contributions from secondary Drell-Yan emission from baryon-anti-baryon,
meson-meson and meson-baryon pairs are not considered in this comparison.
The estimate of ALICE's upper limit of $T_{IMR}$ is
obtained~\cite{Mynote2}
from ALICE data given in Refs.~\cite{ALICE24,ALICE25}.

The horizontal band in Fig.~\ref{fig1} shows the 
constraint~\cite{Tcc}\nocite{Fra15,Ala24} 
\mbox{$T_c^{\hsp Y\hspace*{-2pt}M}= 293(7)~\textrm{MeV}$}
for the critical temperature of the first-order confinement transition, obtained 
from lattice \textrm{QCD} calculations for pure gauge matter.

One can see that the measured dielectron temperatures $T_{IMR}\simeq 0.3~\textrm{GeV}$,
are indeed constant, within roughly a~10 \% uncertainty, in spite of the tremendous increase 
in collision energies by more than two orders of magnitude, 
producing  a strong rise of initial energy densities.

One may wonder, whether this behavior of
intermediate-mass dilepton temperatures at high bombarding energies is caused by
the presence of the early, gluon-dominated phase in heavy-ion collisions
conjectured in previous works \cite{Stoecker:2015nka,Stoecker:2015zea,Vovchenko:2016ijt}.

\section*{Conclusions}

Dilepton data measured at $\sqrt{s_{NN}}$ from
$27$ to~$200~\textrm{GeV}$ (and to $5.02~\textrm{TeV}$) yield a nearly constant dilepton
temperature $T\simeq 290~\textrm{MeV}$ in
the intermediate mass region across a broad range of bombarding
energies. Is this due to the initial lack of light quarks
which precludes the production of dileptons at higher, early
temperatures ($T\gtrsim 600~\textrm{MeV}$), or is this a signature
of a long-lived Yang-Mills mixed phase at  
$T\sim T_c^{\hsp Y\hspace*{-2pt}M}$?
An interesting experimental option is to study IMR dilepton production
in nuclear collisions with light nuclei, like O + O, where
quark suppression effects remain
strong~\cite{Gor26} until the end of a~deconfined  Yang-Mills phase.

\section*{Acknowledgements}

We thank H. Appelsh\"auser, S. Damjanovic, C. Gale, F. Giacosa, J. Harris, 
H. van Hees, L.~McLerran,
M. Panero, and J. Stroth for stimulating discussions. L.M.S. thanks the
Institut f\"ur Theoretische Physik at J.W. Goethe Universit\"at for
financial support.
V.V.~has been supported by the U.S. Department of Energy, 
Office of Science, Office of Nuclear Physics, Early Career Research 
Program under Award Number DE-SC0026065.


\begin{thebibliography}{99}
\bibitem{Mynote2}
    At the LHC energy \mbox{$\sqrt{s_{NN}}=5.02~\textrm{TeV}$} ALICE published
    the IMR dilepton yields~\cite{ALICE24,ALICE25}
    obtained after subtracting contributions of in-medium
    modified heavy-flavor meson decays in the interval
    $1.02~\textrm{GeV} < M < 2.5~\textrm{GeV}$. Only two data boxes with large
    uncertainties and six data points with upper limits are published
    which are compatible with the constraint
    $T_{IMR}<310~\textrm{MeV}$.
\end{thebibliography}
\end{document}